\documentclass{article}
\usepackage{amscd,amsmath,amssymb,bbm}
\usepackage{amsfonts}
\newcommand{\p}[1]{(\ref{#1})}

\newcommand{\cN}{{\cal N}}

\newcommand{\bD}{{\overline{\strut D}}{}}

\newcommand{\be}{\begin{equation}}
\newcommand{\ee}{\end{equation}}
\newcommand{\bea}{\begin{eqnarray}}
\newcommand{\eea}{\end{eqnarray}}
\newcommand{\ba}{\begin{array}}
\newcommand{\ea}{\end{array}}

\newcommand{\nn}{\nonumber}
\newcommand{\NN}{{\mathbb{N}}}
\newcommand{\unity}{\mathbbm{1}}

\def\sfrac#1#2{{\textstyle\frac{#1}{#2}}}
\def\pa{\partial}
\def\di{{\rm d}}
\def\={\ =\ }
\def\ph#1{\phantom{#1}}
\def\vp{\varphi}
\def\vt{\vartheta}
\def\bw{\overline{w}}
\def\bwo{\overline{w}_1^{\phantom{\prime}}}
\def\wt{w_2^{\phantom{\prime}}}
\def\tf{\widetilde{f}}
\def\tg{\widetilde{g}}
\def\th{\widetilde{h}}
\def\tR{\widetilde{R}}

\def\ay{\alpha{\cdot}y}

\begin{document}
\thispagestyle{empty}
\vspace{2cm}
\begin{flushright}
ITP--UH--23/08
\end{flushright}\vspace{2cm}
\begin{center}
{\Large\bf N=4 superconformal n-particle mechanics via superspace}
\end{center}
\vspace{1cm}

\begin{center}
{\large
Sergey Krivonos${}^{a}$, \
Olaf Lechtenfeld${}^{b}$ \ and \
Kirill Polovnikov${}^{b,c}$
}
\end{center}

\begin{center}
${}^a$
{\it Bogoliubov  Laboratory of Theoretical Physics, 
JINR, 141980 Dubna, Russia} 
\vspace{0.2cm}

${}^b$
{\it  Institut f\"ur Theoretische Physik, Leibniz Universit\"at Hannover,
30167 Hannover, Germany} 
\vspace{0.2cm}

${}^c$
{\it  Laboratory of Mathematical Physics, Tomsk Polytechnic University,
634050 Tomsk, Russia}
\end{center}
\vspace{3cm}

\begin{abstract}
\noindent
We revisit the (untwisted) superfield approach to one-dimensional 
multi-particle systems with ${\cal N}{=}4$ superconformal invariance.
The requirement of a standard (flat) bosonic kinetic energy implies
the existence of inertial (super-)coordinates, which is nontrivial 
beyond three particles. We formulate the corresponding integrability
conditions, whose solution directly yields the superpotential, the two
prepotentials and the bosonic potential. The structure equations for
the two prepotentials, including the WDVV equation, follow automatically.
The general solution for translation-invariant three-particle models
is presented and illustrated with examples. For the four-particle case,
we take advantage of known WDVV solutions to construct a $D_3$ and 
a $B_3$ model, thus overcoming a previously-found barrier regarding
the bosonic potential. The general solution and classification remain
a challenge.
\end{abstract}

\newpage
\setcounter{page}{1}

\setcounter{equation}0
\section{Introduction and Summary}
Although conformal multi-particle quantum mechanics (in one space dimension)
is a subject with a long and rich history, its $\cN{=}4$ superconformal
extension has been achieved only recently
\cite{tolik,wyl,bgk,bgl,glp1,glp2,bks,fil}.
Enlarging the conformal algebra~$su(1,1)$ to~$su(1,1|2)$ (with central charge)
imposes severe constraints on the particle interactions, which are not easily
solved.
Firstly, there is a nonzero prepotential~$F$ which must obey a quadratic
homogeneous differential equation of third order known as the
Witten-Dijkgraaf-Verlinde-Verlinde (WDVV) equation~\cite{w,dvv}.
The general solution to the WDVV~equation is unknown, but various classes
of solutions, based on (deformed) Coxeter root systems, have been found
\cite{magra,ves,chaves,feives,glp2}.
Secondly, a second prepotential~$U$ is subject to a linear homogeneous
differential equation of second order, in a given $F$~background.
With the known $F$~solutions, a nonzero~$U$ (needed for nonzero central charge)
has been constructed only for up to three particles,\footnote{
more precisely, translation-invariant irreducible systems of two coupled
relative coordinates plus the decoupled center-of-mass coordinate}
where the WDVV~equation on~$F$ is still empty.
The general $U$~solution is known only for the highly symmetric cases based
on the dihedral root systems~$I_2(p)$, where it depends on three parameters
\cite{glp2,bks}.
The bosonic potential~$V_B$ of the multi-particle system is readily computed
from $U$ and~$F$.
However, beyond three particles, where the WDVV~equation is effective,
the standard logarithmic ansatz for~$U$ is insufficient and must be
enriched by a suitably chosen homogeneous function of degree zero.

Since supersymmetry relates the two prepotentials $F$ and~$U$
to a superpotential~$G$, a superspace approach to these systems
should simplify the analysis. This is what we present in the current paper.
We profit from the fact that $\cN{=}4$, $d{=}1$ superspace is well developed
\cite{leva,ikl1,DI,IvOlaf,BellK}.
Our goal is to construct an ${\cal N}{=}4$ superconformally invariant 
one-dimensional system for $n$~bosonic and $4n$~fermionic physical components. 
Hence we need $n$ copies of an ${\cal N}{=}4$ superfield of type
({\bf1},{\bf4},{\bf3}), containing 1 physical bosonic, 4 fermionic and
3 auxiliary bosonic degrees of freedom. Such a supermultiplet 
is known for a long time~\cite{leva}, and its general action in 
superfields as well as in components was constructed in~\cite{tolik}. 
Everything in this action depends on a single bosonic function~$G$ of the
$n$~superfields, namely the superpotential.
In this paper we investigate the situation 
where two additional properties hold:
\\[-18pt]
\begin{itemize}
\addtolength{\itemsep}{-8pt}
\item the system is superconformally invariant,
\item the bosonic kinetic term is of standard (flat) form 
in suitable coordinates.
\end{itemize}
\vspace{-6pt}
The first condition can rather easily be satisfied.
Contrary to naive expectations, however, the second property implies 
rather intricate constraints on the superpotential (see equation~\p{c6} below), 
which are not solvable in general.
To overcome these problems (at least partially) and to find explicit examples
of ${\cal N}{=}4$ $n$-particle systems with the desired properties, 
we developed an approach whose main features are summarized as follows.

We start from the most general $\cN{=}4$ supersymmetric action for $n$
untwisted ({\bf1},{\bf4},{\bf3}) superfields~$u^A(t)$. Imposing our two 
properties turns out to be equivalent to the existence of `inertial
super-coordinates'~$y^i(t)$ (the $n$ particle locations) together with
integrability and homogeneity conditions on our superfields~$u^A$ as functions
of the~$y^i$. From these conditions we derive the existence and all properties
of the two prepotentials $F(y)$ and~$U(y)$, including the WDVV~equation!
What is more, an explicit construction for~$U$, the third derivatives of~$F$
and the superpotential~$G$ is found in terms of the quadratic homogeneous
functions~$u^A(y)$. Since the homogeneity requirement is easy to fulfil,
the only nontrivial task is to solve the integrability condition on $u^A(y)$
(equation (\ref{gw}) or (\ref{simple}) below).

To obtain explicit solutions, the low-dimensional cases of $n=2,3,4$ are
investigated in detail. Here, we must distinguish between translation
non-invariant irreducible systems and translation invariant reducible systems
of $n$~particles. Any latter one may be constructed from a former one (with
$n{-}1$ `particles') by embedding the former into one dimension higher and
orthogonally to the center-of-mass coordinate to be added.
For $n{=}2$, we reduce the problem to solving certain ordinary differential
equations, which is then done for the dihedral systems and for three examples
including the $A_2$ Calogero case. Their lift to translation-invariant
three-particle models is presented very explicitly. For $n{=}3$
we encounter a system of {\it partial\/} differential equations,
which we cannot solve in general. For a known prepotential~$F$, however, 
the problem simplifies somewhat (equation~(\ref{uwfromF}) below),
and we manage to find the explicit form of $u^A(y)$ for two models based
on the $D_3\simeq A_3$ and $B_3$ root systems. The ensueing prepotentials~$U$
and bosonic potentials~$V_B$ are new in the literature and, for the first time,
overcome the $n{=}3$ barrier of~\cite{glp2}. Finally, we outline how 
to construct the corresponding translation invariant $n{=}4$ models.

\setcounter{equation}0
\section{$\cN{=}4$ supersymmetric $n$-particle systems}
In $\cN{=}4$ superspace one may define two sets of~$\cN{=}4$ superfields
with one physical boson among the components, namely $u^A$ and~$\phi^A$,
restricted by the constraints
\be\label{multiplets}
\bigl[D^a,\,\bD_a\bigr] u^A\=0 \qquad (a) \qquad\textrm{and}\qquad
\bigl( D^a\bD{}^b +D^b\bD{}^a \bigr)\,\phi^A \=0 \qquad (b)\ .
\ee
These constraints define standard and twisted~$\cN{=}4$ supermultiplets,
respectively~\cite{leva,ikl1}.
{}From~(\ref{multiplets}a) it immediately follows that
\be\label{c1}
\partial_t D^2 u^A \= \partial_t \bD{}^2 u^A \=0\ ,
\ee
where \ $D^2=D^aD_a$ \ and \ $\bD^2=\bD_a \bD^a$.
Clearly, these equations result in the conditions
\be\label{c2}
D^2 u^A\=im^A \qquad\textrm{and}\qquad \bD{}^2 u^A \=-im^A\ ,
\ee
where $m^A$ is an arbitrary set of constants.

Considering only standard (nontwisted) superfields for the time being,
the most general $\cN{=}4$ supersymmetric action reads
\be\label{c3}
S\=-\int\!\di t\,\di^2\theta\,\di^2\bar\theta \ G(u^A)
\qquad\textrm{with}\quad A=1,\ldots, n\ ,
\ee
where $G(u^A)$ is an arbitrary function of a set of superfields $u^A$
subject to~\p{c2}.

The bosonic part of the action \p{c3} has the very simple form
(we use the same notation for superfields and their leading components)
\be\label{c4}
S_B\=\sfrac12\int\!\di t\ 
\left[ G_{AB}\,\partial_t u^A \partial_t u^B - G_{AB}\,m^A m^B\right]
\ee
with the evident notation
\be\label{c5}
G_A\equiv\frac{\partial G}{\partial u^A} \ ,\quad
G_{AB}\equiv\frac{\partial^2 G}{\partial u^A \partial u^B} \ ,\quad
G_{ABC}\equiv\frac{\partial^3}{\partial u^A\partial u^B\partial u^C}G\ ,\quad
\textrm{etc.}\ .
\ee
This action has firstly been analyzed in~\cite{tolik}.

We are interested in the subset of actions \p{c3} which features 
superconformal invariance and a {\it flat\/} kinetic term for the bosonic 
variables. The second requirement means that the
Riemann tensor for the metric $G_{AB}$ has to vanish.
One may check that this condition results in the equations
\be\label{c6}
G_{ABX} G^{XY} G_{YCD} - G_{ACX} G^{XY} G_{YBD} \=0
\qquad\textrm{with}\quad G^{XY}G_{YZ} = \delta^X_{\ Z}\ .
\ee
It is not clear how to find the solutions to this equation in full generality.

\setcounter{equation}0
\section{Imposing $\cN{=}4$ superconformal symmetry}\label{sectCONF}
In one dimension the most general superconformal group
is~$D(2,1;\alpha)$~\cite{sorba}. Here we restrict
our consideration to the special case of~$\alpha=-1$,
which corresponds to~SU$(1,1|2)$ symmetry. 
The main reason for this is our wish to retain the potential term
in~\p{c4} with nonzero parameters~$m^A$.
The presence of these constants in the defining superfield constraints~\p{c2}
fixes the scale weight of our superfields~$u^A$ under dilatation 
(the auxiliary components of~$u^A$ must have zero weight). 
This reduces the full superconformal group $D(2,1;\alpha)$ to~SU$(1,1|2)$.

The superconformal group SU$(1,1|2)$ has natural realization 
in~$\cN{=}4$,~$d{=}1$ superspace~\cite{leva} via
\be\label{sc1}
\delta t\=E-\sfrac12\theta^a D_a E -\sfrac12\bar\theta_a\bD{}^a E\ ,\qquad 
\delta\theta^a\=-\sfrac{i}{2} \bD{}^a E\ ,\qquad
\delta\bar\theta_a\=-\sfrac{i}{2} D_a E\ ,
\ee
where the superfunction~$E(t,\theta,\bar\theta)$ 
collects all~SU$(1,1|2)$ parameters:
\be\label{E}
E \= f - 2i (\varepsilon\bar\theta-\theta\bar\varepsilon) + 
\theta^a\bar\theta{}^b B_{(ab)} + 
2(\pa_t\varepsilon\bar\theta+\theta\pa_t{\bar\varepsilon})(\theta\bar\theta) +
\sfrac12(\theta\bar\theta)^2 \pa^2_t{f}
\ee
for
\be
f=a+bt+ct^2 \qquad\textrm{and}\qquad \varepsilon^a = \epsilon^a + t \eta^a\ .
\ee
The bosonic parameters~$a,b,c$ and~$B_{(ab)}$ correspond to translations,
dilatations, conformal boosts and rigid~SU(2) rotations,
while the fermionic parameters~$\epsilon^a$ and~$\eta^a$ correspond to 
Poincar\'e and conformal supersymmetries, respectively.

It is important that by definition the function~$E$ obeys the conditions
\be\label{conf0}
D^2 E \= \bD{}^2E \= \bigl[D^a,\bD_a\bigr]E\=0 \qquad\textrm{and}\qquad 
\partial^3_t E \= \partial^2_t D^a E \= \partial_t D^{(a}\bD{}^{b)}E \=0\ .
\ee
Keeping in mind the transformation
properties of the covariant spinor derivatives~$D^a$ and $\bD_a$,
\be
\delta D^a \= -\sfrac{i}{2}\bigl(D^a \bD_b E\bigr)D^b \qquad\textrm{and}\qquad
\delta \bD_a \= -\sfrac{i}{2}\bigl(\bD_a D^b E\bigr) \bD_b\ ,
\ee
one may check that the constraints~\p{multiplets} are invariant under
the~$\cN{=}4$ superconformal group if the superfields~$u^A$ and $\phi^A$
transform like
\be\label{sc4}
\delta u^A \= \partial_t E\;u^A \qquad\textrm{and}\qquad \delta \phi^A\=0\ ,
\ee
respectively.
Thus, the superfields~$\phi^A$ are superconformal scalars
while the~$u^A$ are vectors.

It is our goal to construct superconformally invariant actions for a set of
$n$ supermultiplets~$u^A$ with $A=1,\ldots,n$.
The variation of the general action \p{c3} under the
superconformal transformations \p{sc1} and \p{sc4} takes the form
\be\label{conf2}
\delta S\=\int\!\di t\,\di^2\theta\,\di^2\bar\theta\ \pa_t{E}\;(-G+u^A\,G_A)\ ,
\ee
which is nullified by the condition
\be\label{conf3}
u^A\,G_A - G \= a_A\,u^A\ ,
\ee
where $a_A$ is an arbitrary set of constants.
The right-hand side disappears after integration
over superspace due to the constraints \p{conf0} and \p{multiplets}.

\setcounter{equation}0
\section{Inertial coordinates}

Since we require the metric $G_{AB}(u)$ to be flat,
there must exist inertial coordinates~$y^i$,
in which the flatness~\p{c6} becomes trivial
because the bosonic action takes the form
\be\label{euclidean}
S_B = \int\!\di t \
\left[ \sfrac12\delta_{ij} \pa_t y^i \pa_t y^j - V_B(y) \right]\ .
\ee
The price for this is a more complicated bosonic potential~$V_B$.
Therefore, we can construct models of the required type by finding
the transformation $u^A=u^A(y)$ to inertial coordinates, with Jacobian
\be\label{53}
u^A_{\ i} \equiv \frac{\partial u^A}{\partial y^i}(y)
\qquad\textrm{and inverse}\qquad
\frac{\partial y^i}{\partial u^A}\bigl(u(y)\bigr)=\left(u^{-1}\right)^i_{\ A} .
\ee

After transforming to the $y$-frame,
the superconformal transformations \p{sc4} become nonlinear,
\be\label{conftr}
\delta y^i = \left( u^{-1} \right)^i_{\ A} u^A \ \partial_t E\,.
\ee
However, the action~\p{euclidean} is invariant only when the
transformation law is
$\delta y^i = \frac{1}{2}\, y^i \partial_t E$.
This demand restricts the variable transformation by
\be\label{ga}
\left( u^{-1} \right)^i_{\ A} u^A = \sfrac{1}{2}\, y^i
\qquad \rightarrow \qquad y^i\,u^A_{\ i} = 2\,u^A.
\ee
Hence, superconformal invariance requires $u^A$ to be a homogeneous quadratic
function of the~$y^i$.

A rigid SU(2) rotation brings the constraints \p{multiplets} and \p{c2}
into the equivalent form
\be\label{52}
D^2 u^A = 0,\quad \bD{}^2 u^A =0,\quad \bigl[ D^a, \bD_a \bigr] u^A =2\,m^A .
\ee
In the new coordinates, they become
\bea\label{54}
&& u^A_{\ i} D^2 y^i+ (\pa_i u^A_{\ j}) D^a y^i D_a y^j = 0,\nn \\[6pt]
&& u^A_{\ i} \bD{}^2 y^i+ (\pa_i u^A_{\ j}) \bD_a y^i \bD^a y^j = 0, \\[6pt]
&& u^A_{\ i} \bigl[ D^a,\bD_a\bigr] y^i + 2(\pa_i u^A_{\ j})D^a y^i\bD_a y^j
= 2\,m^A \,, \nn
\eea
which we rewrite as~\cite{wyl}
\bea\label{56}
&& D^2 y^i- f^i_{\ kj} D^a y^k D_a y^j=0, \nn \\[6pt]
&& \bD^2 y^i- f^i_{\ kj} \bD_a y^k \bD^a y^j=0, \\[6pt]
&& \bigl[D^a,\bD_a\bigr]y^i - 2 f^i_{\ kj} D^a y^k\bD_a y^j + 2\,U^i =0\,,\nn
\eea
after introducing a flat connection and a covariantly constant vector via
\be\label{f}
f^i_{\ kj}=
-\left(u^{-1}\partial_k u\right)^i_{\ j} =
-\left(u^{-1}\right)^i_{\ A} u^A_{\ kj}
\qquad\textrm{and}\qquad
\quad U^i = -\left( u^{-1} \right)^i_{\ A} m^A \,,
\ee
in obvious notation.
By construction, the integrability conditions of the system~\p{56},
\bea
&& \partial_{[k} f^i_{\ m]n} - f^i_{\ j[k}f^j_{\ m]n}=0 , \label{59} \\
&& \partial_j U^i -f^i_{\ jk}U^k =0 , \label{510}
\eea
are automatically satisfied.
No restriction (besides invertibility)
on the matrix $(u^A_{\ i})$ appears.

Let us come back to the superfield action \p{c3} and consider
the superpotential as a function of the inertial coordinates,
writing again $G(y)$ in place of $G(u(y))$ in a slight abuse of notation,
so that $G_i$, $G_{ij}$ etc.\ denote its derivatives with respect to~$y$.
After integration over the $\theta$s in \p{c3} and using the constraints \p{56}
we arrive at
\be\label{actA}
S_B=-\sfrac12 \int\!\di t \ \Bigl[
\Bigl( G_{ij}+ G_k f^k_{\ ij}\Bigr) \pa_t{y}^i \pa_t{y}^j -
\pa_k\Bigl(G_i\ \bigl(u^{-1}\bigr)^i_{\ A}\Bigr)\bigl(u^{-1}\bigr)^k_{\ B}
m^A m^B \Bigr] ,
\ee
which may also be obtained by directly subjecting \p{c4}
to the change of variables.
Comparing with the defining property \p{euclidean}, we read off that
\be\label{sys1}
G_{ij}+G_k f^k_{\ ij}=-\delta_{ij} ,
\ee
which simplifies the potential term to\footnote{
This is the classical bosonic potential. After quantization, it picks up
an additional contribution of $\sfrac18\hbar^2 f_{ijk}f_{ijk}$.}
\be\label{pot}
V_B=
\sfrac12\delta_{ij}\left(u^{-1}\right)^i_{\ A}\left(u^{-1}\right)^j_{\ B}m^Am^B
= \sfrac12 \delta_{ij} U^i U^j .
\ee
Differentiating the condition \p{sys1} with respect to~$y^m$ we get
\be\label{512}
0 = G_{ijm}+G_{mk}f^k_{\ ij}+G_k \partial_m f^k_{\ ij} =
G_{ijm}- \delta_{mk} f^k_{\ ij} +
G_k\left( \partial_m f^k_{\ ij} - f^k_{\ lm}f^l_{\ ij}\right).
\ee
In view of \p{59}, antisymmetrizing in $i$ and $m$ yields\footnote{
Since the inertial metric is euclidean, we may freely raise and lower
inertial indices.}
\be
\delta_{k[m} f^k_{\ i]j} =0 \qquad\longrightarrow\qquad
\delta_{mk} f^k_{\ ij} \equiv f_{mij} = f_{imj} = f_{ijm},
\ee
so that our flat connection is symmetric in all three indices.

A flat connection as defined in \p{f} can be totally symmetric
(after lowering all indices) if and only if the inverse Jacobian is integrable,
\be\label{sys2}
\frac{\partial y^i}{\partial u^A}\bigl(u(y)\bigr) \equiv
\left( u^{-1}\right)^i_{\ A} =: w_{A,i} =
\partial_i w_A \equiv \frac{\partial w_A}{\partial y^i}(y),
\ee
which establishes the existence of a set of superfields~$w_A$ `dual' to~$u^A$.
The $w_A$ can be shifted by integration constants.
It is instructive to rewrite formulae by replacing $u^{-1}$ by $w$.
Beginning with
\be\label{gw}
w^{\phantom{B}}_{A,i}\,u^B_{\ i} = \delta_A^{\ B}
\qquad \longleftrightarrow \qquad
w^{\phantom{B}}_{A,i}\,u^A_{\ j} = \delta _{ij}
\ee
and the superconformality condition \p{ga},
\be\label{ga2}
y^i\,u^A_{\ i} = 2\,u^A \qquad \longleftrightarrow \qquad
w^{\phantom{B}}_{A,i} u^A = \sfrac12 y^i\ ,
\ee
we introduce the notation $p^{\phantom{B}}_A q^A=p\cdot q$ and
find by repeated differentiation and contraction with $y^k$ that
\bea
&& y^k\,u^A_{\ k} = 2\,u^A \quad \longrightarrow \quad
y^k\,u^A_{\ ki} = u^A_{\ i} \qquad\textrm{and}\qquad
y^k\,w_{A,ki} = -w_{A,i} \quad \longrightarrow \quad
y^k\,w_{A,k} = c_A \ , \label{homog}\\[4pt]
&& 3w_{ij}{\cdot}u = -\sfrac32\delta_{ij} =
w{\cdot}u_{ij} - \pa_i\pa_j(w{\cdot}u) \ ,\\[4pt]
&& f_{ijk}= -w_{ijk}{\cdot}u = w_{ij}{\cdot}u_{k} = -w_{i}{\cdot}u_{jk}
= w{\cdot}u_{ijk} - \pa_i\pa_j\pa_k(w{\cdot}u) \ ,\label{fijk}
\eea
with some constants $c_A$.
In the last line, indices may be permuted freely.
Playing a bit more, one finds
\bea
&& w_{ij}{\cdot}u_{kl} = w_{ij}{\cdot}u_m\,w_m{\cdot}u_{kl}
= w_m{\cdot}u_{ij}\,w_{kl}{\cdot}u_m = w_{kl}{\cdot}u_{ij}\ ,\\[4pt]
&& 2\pa_l f_{ijk} = \pa_l(w_{jk}{\cdot}u_i-w_i{\cdot}u_{jk})
= w_{jkl}{\cdot}u_i - w_i{\cdot}u_{jkl}\ ,
\eea
proving that
\be
\pa_{[l} f_{k]ij} =0 \qquad\longrightarrow\qquad
f_{ijk} = \pa_i\pa_j\pa_k F \qquad\textrm{and}\qquad
f^{[k}_{\ im} f^{l]}_{\ mj } = 0\ .
\ee
Hence, there exists a prepotential $F$ obeying the WDVV equation.

The homogeneity relations~\p{homog} imply that
there exists a `radial coordinate',
\be\label{rad}
c{\cdot}u_k = y^k \qquad\longrightarrow\qquad
c{\cdot}u =\sfrac12 y^ky^k =: \sfrac12 R^2\ .
\ee
In view of this, it is reasonable to choose
\be
u^1 = R^2 \quad\textrm{and}\quad w_1 = \sfrac12 \ln R + \bw_1
\qquad\longrightarrow\qquad
c_1=\sfrac12 \quad\textrm{and}\quad c_{A>1}=0 \ .
\ee
Furthermore, contractions of $f_{ijk}$ simplify,
\be\label{contractf}
w_{A,i}\,f_{ijk} = w_{A,jk}\ ,\quad
u^A_{\ i}\,f_{ijk} =-u^A_{\ jk} \qquad\textrm{and}\qquad
y^i\,f_{ijk} = -\delta_{jk}\ ,
\ee
and the vector~$U_i=\delta_{ij}U^j$ obeys
\be
U_i = -w_{A,i}\,m^A = \pa_i U \qquad\longrightarrow\qquad
U = -w_Am^A \qquad\textrm{and}\qquad
y^i\,U_i = - c_A\,m^A =: -C\ .
\ee
Thus, all the `structure equations' of \cite{glp1} are fulfilled
precisely by
\be\label{prepotentials}
\pa_i\pa_j\pa_k F = f_{ijk} \qquad\textrm{and}\qquad \pa_i U = U_i
\ee
and the central charge~$C$.

With the help of the `dual superfields' $w_A$, one can give a simple
expression for the superpotential $G(y)$, namely
\be\label{F}
G =  - u^A w_A \= -\sfrac12 R^2\ln R - R^2\bw_1 - u^{A>1} w_{A>1}\ .
\ee
Employing the relations above, it is readily verified that this function
indeed obeys~\p{sys1} and thus leads to the bosonic action~\p{euclidean}.
In the inertial coordinates, the superconformality condition~\p{conf3}
acquires the form
\be
y^i G_i - 2 G - 2 a_A u^A = 0 \ .
\ee
The superpotential $G$ given by~\p{F} does satisfy this constraint,
provided the constants $c_A$ and $a_A$ are related as $c_A=-2\,a_A$.
In view of~\p{rad}, this yields the homogeneity relation
\be\label{ghom}
y^i G_i - 2 G + \sfrac{1}{2}y^iy^i = 0 \,.
\ee
Incidentally, the prepotential~$F$ defined in~\p{prepotentials}
respects just the same homogeneity relation, as is found by twice
integrating the last equation in~\p{contractf}.

So, for the construction of $\cN{=}4$ superconformal mechanics models,
in principle one needs to solve only two equations, namely
\p{gw} and \p{ga2}. All other relations and conditions follow from these!
The homogeneity condition~\p{ga2} is easy to satisfy: the $u^A$ must be
homogeneous of degree two as functions of~$y$. Nontrivial, however, is
the integrability condition~\p{gw}. Its derivative may be recast in a
different form:
\be\label{simple}
0 = w_{ik}{\cdot}u_j + w_i{\cdot}u_{jk} = -w_j{\cdot}u_{ik} + w_i{\cdot}u_{jk}
\qquad\longrightarrow\qquad
u^{[A}_{\ i}\,u^{B]}_{\ ij} = 0\ ,
\ee
after contracting with two Jacobians. This equation looks deceivingly simple.
Contracting it with $w_{A,k}w_{B,l}$ we reproduce the total symmetry of
$f_{ijk}=w_i{\cdot}u_{jk}$ and thus the integrability $w_{A,i}=\pa_i w_A$.
If~$F$ is known otherwise, e.g.~from solving the WDVV equation, it is
easier to reconstruct $u^A$ or $w_A$ from~(\ref{contractf}),
\be \label{uwfromF}
u^A_{\ ij}\ +\ f_{ijk}\,u^A_{\ k} \= 0 \qquad\textrm{and}\qquad
w_{A,ij}\ -\ f_{ijk}\,w_{A,k} \= 0\ .
\ee
With $f_{ijk}$ being totally symmetric, 
any one of these equations is equivalent to (\ref{simple}). 
Their advantage is the linearity, which allows superpositions,
as long as we respect~(\ref{rad}).

It is also worthwhile to consider $w_A$ as a function of the $u^A$, i.e.
\be
w_A \= w_A\left(y(u)\right) \qquad\longrightarrow\qquad
\pa_B=w_{B,i}\;\pa_i \quad\textrm{and}\quad \pa_i=u^A_{\ i}\,\pa_A \ .
\ee
Then,
\be
w_{AB} \ \equiv\ \pa_B w_A \= w_{B,i}\,w_{A,i} \= w_{BA} ,
\ee
and the bosonic potential \p{pot} can be rewritten as
\be
V_B \= \sfrac12 m{\cdot}w_i\ m{\cdot}w_i \= \sfrac12 m^A m^B\,w_{AB}\ .
\ee
Furthermore, we can directly reconfirm \p{conf3} and discover that
\be
G_{AB}=-w_{AB} \quad\textrm{and}\quad G_A=-w_A-\sfrac12 c_A
\qquad\longrightarrow\qquad
w_A = -\pa_A(G+\sfrac12 c{\cdot}u) = -\pa_A(G+\sfrac14 u^1)\ .
\ee
We note that, since $w_A$ is only determined up to a constant,
a linear function of~$u^A$ may be added to~$G$, e.g.~to achieve $w_A=-\pa_A G$.
As expected, the superpotential~$G(y)$ determines both $U$ and~$f_{ijk}$,
\be
U \= m^A \pa_A G \qquad\textrm{and}\qquad
G_{ij}+G_k f^k_{\ ij}=-\delta_{ij}\ ,
\ee
albeit rather indirectly.

\setcounter{equation}0
\section{Two-dimensional systems}

In the simplest situation of $n{=}2$, all equations can be solved in principle.
Indeed, the integrability condition~\p{simple} then merely implies that
$u^1$ and~$u^2$ are homogeneous quadratic functions of $y^1$ and $y^2$.
So -- moving the inertial index down for notational simplicity --
let us take
\be \label{uydef}
u^1 = y_1^2 + y_2^2 =: R^2 \qquad\textrm{and}\qquad
u^2 = R^2\,h(\vp) \qquad\textrm{with}\quad \tan\vp=\sfrac{y_2}{y_1}\ .
\ee
With
\be
(u^A_{\ i}) \= \left( \begin{array}{cc}
2y_1 & 2y_2 \\[4pt] 2y_1 h-y_2 h' & 2y_2 h+y_1 h'
\end{array} \right)
\ee
the condition \p{simple} is identically satisfied.
Inversion of this matrix produces
\be
u^{-1} \= \left( \begin{array}{cc}
\sfrac{y_1}{2R^2}{+}\sfrac{y_2h}{R^2h'} & -\sfrac{y_2}{R^2h'} \\[4pt]
\sfrac{y_2}{2R^2}{-}\sfrac{y_1h}{R^2h'} & \phantom{-}\sfrac{y_1}{R^2h'}
\end{array} \right)
\= \left( \begin{array}{cc}
w_{1,1} & w_{2,1} \\[8pt] w_{1,2} & w_{2,2}
\end{array} \right)\ ,
\ee
and we read off that
\be
w_1 \= \sfrac12\ln R + \bwo(\vp)  \quad\textrm{and}\quad
w_2 \= \wt(\vp) \qquad\textrm{with}\qquad
\bw'_1=-\sfrac{h}{h'} \quad\textrm{and}\quad w'_2=\sfrac{1}{h'}\ .
\ee
This yields the superpotential
\be
G \= -\sfrac12 R^2\ln R - R^2\,\tg(\vp)
\qquad\textrm{with}\qquad
\tg \= \bwo+h\,\wt 
\quad\longrightarrow\quad
\tg' \= h'\wt \= \sfrac{\wt}{w_2'}\ .
\ee

Let us make a matching ansatz for the WDVV prepotential,
\be
F \= -\sfrac12 R^2\ln R - R^2\,\tf(\vp)\ .
\ee
Then, from $\pa_i\pa_j\pa_kF=-w_i{\cdot}u_{jk}$ we learn that
\be \label{fhw}
\tf^{\prime\prime\prime}+4\tf' \= \sfrac{h''}{h'}
\= -\sfrac{w''_2}{w'_2} \= -\sfrac{\bw''_1+1}{\bw'_1}
\qquad\longleftrightarrow\qquad
h'\ \propto\ {\rm e}^{\tf''+4\tf} \ ,
\ee
and the bosonic potential reads
\be
V_B \= (\sfrac{m^1}{2\,R})^2 + (\sfrac{m^1h-m^2}{h'\,R})^2
\qquad\textrm{with}\quad C=\sfrac12 m^1\ .
\ee

\subsection{Dihedral systems}

A highly symmetric class of models is based on the dihedral
root systems~$I_2(p)$ for $p\in\NN$,
\be
\alpha\cdot y \= \cos(k\pi/p)\,y_1\ +\ \sin(k\pi/p)\,y_2
\qquad\textrm{for}\quad k=0,1,\ldots,p{-}1\ .
\ee
The WDVV prepotential for these systems was found to be~\cite{glp2}
\be
F \= -\sfrac12 f_R\,R^2\ln R\ -\ \sfrac12 f_p \!\!\sum_{\alpha\in I_2(p)}\!\!
(\ay)^2 \ln|\ay| \qquad\textrm{with}\quad f_R+\sfrac{p}{2}\,f_p =1\ ,
\ee
which corresponds to
\be
\tf(\vp) \= \sfrac12 f_p \ \sum_{k=0}^{p-1}
\cos^2(\vp{-}\sfrac{k\pi}{p})\ \ln|\cos(\vp{-}\sfrac{k\pi}{p})|\ .
\ee
Differentiating, we obtain (modulo irrelevant integration constants)
\be
\ln h' \= \tf''+4\tf \= \sfrac{p}{2} f_p + f_p
\sum_{k=0}^{p-1} \ln|\cos(\vp{-}\sfrac{k\pi}{p})|
\qquad\longrightarrow\qquad
h' \ \propto\ \bigl[\sin(p\vp{+}p\sfrac{\pi}{2})\bigr]^{f_p}
\ee
and thus
\be
h(\vp) \= h_0\,\cos(p\vp{+}p\sfrac{\pi}{2})\ {}_2F_1\bigl(
\sfrac12,\sfrac{1-f_p}{2},\sfrac32,\cos^2(p\vp{+}p\sfrac{\pi}{2})\bigr)\ .
\ee
This result simplifies for\\[-12pt]
\begin{itemize}
\addtolength{\itemsep}{-4pt}
\item $f_p=+1:\quad h(\vp) = h_0\,\cos(p\vp{+}p\sfrac{\pi}{2})
\qquad\qquad\qquad\bullet\
f_p=-1:\quad h(\vp) = h_0\,\ln\tan(p\vp{+}p\sfrac{\pi}{2})$
\item $f_p=-2:\quad h(\vp) = h_0\,\cot(p\vp{+}p\sfrac{\pi}{2})
\qquad\qquad\qquad\bullet\
f_p=\ \ \,0:\quad h(\vp)= h_0\,\vp$
\end{itemize}
for which $w_A$ and $G$ are readily computed.
With impunity the roots may be rotated by a common angle~$\delta$,
which corresponds to $\vp\to\vp{-}\delta$ in all equations.
We provide three simple examples.

\subsection{First example}

An easy choice is
\be
h(\vp) = \sin 2\vp \qquad\leftrightarrow\qquad
u^2 = 2\,y_1 y_2 \qquad\textrm{so that}\qquad
u^1{\pm}u^2=(y_1{\pm}y_2)^2\ ,
\ee
which leads to
\be\label{ex1}
u^{-1} \= \left( \begin{array}{cc}
\frac{y_1}{2(y_1^2-y_2^2)} & -\frac{y_2}{2(y_1^2-y_2^2)} \\
-\frac{y_2}{2(y_1^2-y_2^2)} & \frac{y_1}{2(y_1^2-y_2^2)}
\end{array} \right)
\= \left( \begin{array}{cc}
w_{1,1} & w_{2,1} \\[8pt] w_{1,2} & w_{2,2}
\end{array} \right) \ .
\ee
This integrates to
\be
w_A \= \sfrac14\ln|y_1{+}y_2|\pm\sfrac14\ln|y_1{-}y_2|
\= \sfrac18\ln|u^1{+}u^2|\pm\sfrac18\ln|u^1{-}u^2|
\ee
with the upper (lower) sign corresponding to $A{=}1$ ($A{=}2$),
and further yields the superpotential
\be
\begin{aligned}
G \= &-\sfrac18(u^1{+}u^2)\ln|u^1{+}u^2| -\sfrac18(u^1{-}u^2)\ln|u^1{-}u^2| \\
\= &-\sfrac14(y_1{+}y_2)^2\ln|y_1{+}y_2| -\sfrac14(y_1{-}y_2)^2\ln|y_1{-}y_2|
\= F
\end{aligned}
\ee
as well as $(c_1,c_2)=(\sfrac12,0)$. It is obvious that $G=-u^Aw_A$.
Depending on the value of $(m^1,m^2)$,
the bosonic potential is a linear combination of $(y_1{+}y_2)^{-2}$ and
$(y_1{-}y_2)^{-2}$.
We recognize the roots of~$D_2$ here. A rotation by $\delta{=}\sfrac{\pi}4$
produces the (decoupled) $I_2(2)=A_1{\oplus}A_1$ system with $h=\cos 2\vp$
as well as $f_p{=}1$ and $f_R{=}0$.
The decoupling of the center of mass
$u^1{+}u^2=(y_1{+}y_2)^2$ renders this example a bit trivial.

\subsection{Second example}

For a more complicated case, consider
\be\label{ex11}
h(\vp) = \sqrt{\sin^4\vp+\cos^4\vp} \qquad\leftrightarrow\qquad
u^2 = \sqrt{y_1^4+y_2^4} \qquad\textrm{so that}\qquad
(u^1)^2-(u^2)^2\=2 y_1^2\,y_2^2\ .
\ee
The matrix $u^{-1}$ may easily be found to be
\be\label{ex12}
u^{-1} \= \left( \begin{array}{cc}
-\frac{y_2^2}{2y_1(y_1^2-y_2^2)} & 
\frac{\sqrt{y_1^4+y_2^4}}{2y_1(y_1^2-y_2^2)}\\
\frac{y_1^2}{2y_2(y_1^2-y_2^2)} & 
-\frac{\sqrt{y_1^4+y_2^4}}{2y_2(y_1^2-y_2^2)}
\end{array} \right)
\= \left( \begin{array}{cc}
w_{1,1} & w_{2,1} \\[8pt] w_{1,2} & w_{2,2}
\end{array} \right) \ ,
\ee
which can be integrated to
\be \label{ex14}
\begin{aligned}
w_1 \=&
\sfrac14\ln{y_1^2}+\sfrac14\ln{y_2^2}
-\sfrac14\ln{|y_1^2{-}y_2^2|}\ , \\[4pt]
w_2 \=&
-\sfrac14\ln{y_1^2}-\sfrac14\ln{y_2^2}
+\sfrac14\ln{\bigl(y_1^2{+}\sqrt{y_1^4{+}y_2^4}\,\bigr)}
+\sfrac14\ln{\bigl(y_2^2{+}\sqrt{y_1^4{+}y_2^4}\,\bigr)} \\
& +\sfrac{1}{2\sqrt{2}}\ln{|y_1^2{-}y_2^2|}
-\sfrac{1}{2\sqrt{2}}
\ln{\bigl(y_1^2{+}y_2^2{+}\sqrt{2}\sqrt{y_1^4{+}y_2^4}\,\bigr)}\ .
\end{aligned}
\ee
It is amusing to check that indeed
\be\label{ex15}
\left(\pa_i w_A \right) u^A = \sfrac12 y_i \quad,\qquad
\left(\partial^2_{ij}w_A\right) u^A = -\sfrac12\delta_{ij}
\qquad\textrm{and}\qquad G_A = -w_A-\sfrac12 c_A\ ,
\ee
as it must be by construction.
The simplest form of the bosonic potential \p{pot} occurs for the choice
$(m^1,m^2)=(0,m)$, namely
\be\label{ex13}
V_B\ \sim\ 
\frac1{y_1^2}+\frac1{y_2^2}+\frac1{(y_1{-}y_2)^2}+\frac1{(y_1{+}y_2)^2}\ .
\ee
We recognize the roots of the $I_2(4)=BC_2$ system. This is not surprising,
since $h=\sfrac12\sqrt{\cos 4\vp+3}$ is a simple deformation of the
dihedral construction.
This model is not translation invariant. In fact, the only two-dimensional
model with this property is our first example above.

\subsection{Third example}

Finally, let us present the standard Calogero example based on the
$A_2$ root system,
\be \label{a2h}
h(\vp) = \sin 3\vp \qquad\leftrightarrow\qquad
u^2 = \sfrac{3y_1^2y_2^{\ph{2}}-y_2^3}{\sqrt{y_1^2+y_2^2}} \ .
\ee
It leads to
\be
u^{-1} \= \left( \begin{array}{cc}
\sfrac{9}{y_1}+\sfrac{2\,y_1}{9(y_1^2-3y_2^2)}+\sfrac{y_1}{6(y_1^2+y_2^2)} &
\sfrac{y_2\,\sqrt{y_1^2+y_2^2}}{3y_1(3y_2^1-y_1^2)} \\
-\sfrac{2\,y_2}{y_1^2-3y_2^2}+\sfrac{y_2}{6(y_1^2+y_2^2)} &
\sfrac{\sqrt{y_1^2+y_2^2}}{3(y_1^2-3y_2^2)}
\end{array} \right)
\= \left( \begin{array}{cc}
w_{1,1} & w_{2,1} \\[8pt] w_{1,2} & w_{2,2}
\end{array} \right) \ ,
\ee
which produces
\be \label{wthird}
\begin{aligned}
w_1 &\ =\sfrac19\ln\left| y_1\,(y_1{-}\sqrt{3}y_2)(y_1{+}\sqrt{3}y_2)\right|
+ \sfrac16\ln R \=
\sfrac{1}{18} \ln|u^1{+}u^2| + \sfrac{1}{18} \ln|u^1{-}u^2|
+  \sfrac{5}{36} \ln |u^1| \ , \\[4pt]
w_2 &\= \sfrac19\ln\left|
\sfrac{y_1\,(y_2-\sqrt{3}y_1+2R)(y_2+\sqrt{3}y_1+2R)}
{(y_1+\sqrt{3}y_2)(y_1-\sqrt{3}y_2)(y_2+R)}\right| \=
\sfrac{1}{18} \ln|u^1{+}u^2|- \sfrac{1}{18} \ln|u^1{-}u^2|
\end{aligned}
\ee
and the superpotential
\be \label{Gthird}
G \=
\sfrac{1}{18} \left(u^1{+}u^2\right) \ln |u^1{+}u^2| +
\sfrac{1}{18} \left(u^1{-}u^2\right) \ln |u^1{-}u^2| +
\sfrac{5}{36} u^1 \ln|u^1|\ .
\ee
For the bosonic potential $V_B$, please proceed to the following section.

\setcounter{equation}0
\section{Embedding into three dimensions}

The generic two-dimensional system is irreducible and thus not translation
invariant. To generate translation-invariant models, we may take the
inertial $y$~coordinates as relative coordinates in a three-particle system,
whose absolute coordinates $(x^1,x^2,x^3)$ comprise the center-of-mass
combination
\be
u^0\=(x^1{+}x^2{+}x^3)^2 \qquad\textrm{while}\qquad
u^1 = u^1(x^{\mu\nu}) \ ,\quad u^2 = u^2(x^{\mu\nu})
\qquad\textrm{with}\quad x^{\mu\nu}:=x^\mu{-}x^\nu
\ee
live in the `relative-motion plane' orthogonal to the center-of-mass motion.
Our notation reflects the $3=1{+}2$ split of this reducible system.
To find the relation between the 3d~coordinates $x^\mu$ and the 2d~coordinates
$y^i$, we have to formulate the embedding map~\cite{glp2},
\be
y^i \= M^i_{\ \mu}\,x^\mu \qquad\textrm{with}\qquad
\bigl(M^i_{\ \mu}\bigr) \= \begin{pmatrix}
\sfrac{1}{\sqrt{2}} &      -\sfrac{1}{\sqrt{2}} & \ph{-}0 \\[6pt]
\sfrac{1}{\sqrt{6}} & \ph{-}\sfrac{1}{\sqrt{6}} & -\sfrac{2}{\sqrt{6}}
\end{pmatrix}\ .
\ee
Here, the matrix $M$ effects a partial isometry,
\be
M\,M^\top \= \unity_2 \qquad\textrm{and}\qquad
M^\top M \= P \= \sfrac13\,\Bigl(\!\!\begin{smallmatrix}
\ph{-}2 & -1 & -1 \\ -1 & \ph{-}2 & -1 \\ -1 & -1 & \ph{-}2
\end{smallmatrix}\Bigr)\ ,
\ee
where $P$ is the projection onto the relative-motion plane.
By a slight abuse of notation, we write $u^A(y{=}Mx)=u^A(x)$ and
embed~(\ref{uydef}),
\be \label{uxdef}
\begin{aligned}
u^1(x) &\= x^\top P\,x \=
\sfrac13\bigl\{(x^{12})^2+(x^{23})^2+(x^{31})^2\bigr\}\ =:\ \tR^2 \ ,\\[4pt]
u^2(x) &\= \tR^2\,h(\vp) \qquad\textrm{with}\qquad
\sin\vp=\sfrac{x^1+x^2-2x^3}{\sqrt{6}\tR} \quad\textrm{and}\quad
\cos\vp=\sfrac{x^1-x^2}{\sqrt{2}\tR}\ .
\end{aligned}
\ee
This will automatically take care of the integrability condition~\p{simple}.
Permutations of the $x^\mu$ are generated by the reflection
$\vp\mapsto\pi{-}\vp$ and a $\sfrac{2\pi}{3}$ rotation in the
relative-motion plane~\cite{glp2}.
Therefore, if we want to describe a system of three identical particles,
the function $h(\vp)$ better be invariant under these actions, for instance
by taking
\be
h(\vp) \= \th(\vp)\,\th(\vp{+}\sfrac{2\pi}{3})\,\th(\vp{-}\sfrac{2\pi}{3})
\qquad\textrm{with}\qquad \th(\pi{-}\vp)=\th(\vp)\ .
\ee

The $A_2$ Calogero model arises from the simple choice \
$\th(\vp)=-\root 3\of 4\,\sin\vp$, which gives
\be \label{h1}
h(\vp)\= -4\sin(\vp)\,\sin(\vp{+}\sfrac{2\pi}{3})\,\sin(\vp{-}\sfrac{2\pi}{3})
\= \sin 3\vp \= \sfrac{\sqrt{2}\,
\left(2x^1{-}x^2{-}x^3\right)
\left(2x^2{-}x^3{-}x^1\right)
\left(2x^3{-}x^1{-}x^2\right)}
{\bigl[(x^{12})^2+(x^{13})^2+(x^{23})^2\bigr]^{\frac32}}
\ee
and thus
\be\label{3cal2}
u^2 \= \frac{\sqrt{2}\left(x^{12}{+}x^{13}\right)\left(x^{21}{+}x^{23}\right)
\left(x^{31}{+}x^{32}\right)} {3\,\sqrt{(x^{12})^2+(x^{13})^2+(x^{23})^2 }}\ ,
\ee
which also follows directly from~(\ref{a2h}).
For this choice we
integrate the matrix $(u^{-1})^\mu_{\ A} = \pa_\mu w_A$ to get
\be
\begin{aligned}
w_0&\=\sfrac16 \ln\left|x^1{+}x^2{+}x^3\right|
\= \sfrac{1}{12}\ln u^0\ ,\\[4pt]
w_1&\=\sfrac{1}{9} \ln|x^{12}x^{13}x^{23}| +\sfrac16 \ln\tR
\= \sfrac{1}{18} \ln|u^1{+}u^2| + \sfrac{1}{18} \ln|u^1{-}u^2|
+  \sfrac{5}{36} \ln |u^1|\ , \\[4pt]
w_2&\=\sfrac{1}{9}\ln\left|
\sfrac{x^{21}\ (x^{23}+x^{21}+\sqrt{6}\tR)\ (x^{12}+x^{13}+\sqrt{6}\tR)}
      {x^{23}\ x^{13}\ (x^{32}+x^{31}-\sqrt{6}\tR)}\right|
\= \sfrac{1}{18} \ln|u^1{+}u^2|- \sfrac{1}{18} \ln|u^1{-}u^2| \ .
\end{aligned}
\ee
As expected, $w_1(u)$ and $w_2(u)$ agree with the functions in~(\ref{wthird}).
Beyond the center-of-mass term, the superpotential then reproduces
the result of~(\ref{Gthird}) (see also~\cite{bks}),
\be\label{3cal1}
G\=\sfrac{1}{12} u^0 \ln|u^0| +
\sfrac{1}{18} \left(u^1{+}u^2\right) \ln |u^1{+}u^2| +
\sfrac{1}{18} \left(u^1{-}u^2\right) \ln |u^1{-}u^2| +
\sfrac{5}{36} u^1 \ln|u^1|\ .
\ee
The possible potential terms are specified by a choice of the constants
$m^A$ in the basic constraints on the superfields \p{c2}.
They completely agree with the results of~\cite{glp2} on the $A_2$ model.
The coupling $m^0$ goes with the center of mass.
The general bosonic potential for $m^1 \neq 0 $ and $m^2 \neq 0$
is not very illuminating, so we display two special cases:
\bea\label{3cal4}
{V_B}\lower2pt\hbox{$\big|_{m^1=0}$} &=& \sfrac{1}{81}(m^2)^2 \left(
\frac{1}{(x^{12})^2}+\frac{1}{(x^{23})^2}+\frac{1}{(x^{31})^2}\right)+
\frac{\sfrac{1}{24}(m^0)^2}{(x^1{+}x^2{+}x^3)^2}\ ,\\[4pt]
{V_B}\lower2pt\hbox{$\big|_{m^2=0}$} &=& \sfrac{1}{81}(m^1)^2 \left(
\frac{1}{(x^{12})^2}+\frac{1}{(x^{23})^2}+\frac{1}{(x^{31})^2}\right)+
\frac{\sfrac{5}{24}(m^1)^2}{(x^{12})^2+(x^{23})^2+(x^{31})^2} +
\frac{\sfrac{1}{24}(m^0)^2}{(x^1{+}x^2{+}x^3)^2}\ . \nn\eea

Shifting $\vp$ by a constant should produce an equivalent formulation of the
Calogero model. For instance,
\be \label{h2}
h(\vp) = \cos 3\vp \qquad\longrightarrow\qquad
u^2 \= \frac{\sqrt{6}\ x^{12}\,x^{23}\,x^{31}}
           {\sqrt{(x^{12})^2+(x^{13})^2+(x^{23})^2 }}\ .
\ee
In this case, we find
\bea \nn
&& w_0\=\sfrac16\ln\left|x^1{+}x^2{+}x^3\right|\=\sfrac{1}{12}\ln u^0\ ,\\[4pt]
&& w_1\=
\sfrac19 \ln\left|(x^{13}{+}x^{23})(x^{21}{+}x^{31})(x^{32}{+}x^{12})\right|
+\sfrac16 \ln\tR \=
\sfrac{1}{18} \ln|u^1{+}u^2| + \sfrac{1}{18} \ln|u^1{-}u^2|
+ \sfrac{5}{36} \ln |u^1|\ , \nn\\[4pt]
&& w_2\=\sfrac19 \ln\left|
\sfrac{(x^{21}+x^{23})(x^{12}+\sqrt{2}\tR)(x^{23}+\sqrt{2}\tR)}
      {(x^{31}+x^{32})(x^{12}+x^{13})(x^{31}-\sqrt{2}\tR)}\right| \=
\sfrac{1}{18} \ln|u^1{+}u^2|- \sfrac{1}{18} \ln|u^1{-}u^2| \ ,
\eea
and obtain bosonic potentials
\be\label{3cal5}
\begin{aligned}
{V_B}\lower2pt\hbox{$\big|_{m^1=0}$} &\= \sfrac{1}{27}(m^2)^2 \left(
\frac{1}{(x^{13}{+}x^{23})^2}+\textrm{cyclic}\right)
+ \frac{\sfrac{1}{24}(m^0)^2}{(x^1{+}x^2{+}x^3)^2}\ ,\\[4pt]
{V_B}\lower2pt\hbox{$\big|_{m^2=0}$} &\= \sfrac{1}{27}(m^1)^2 \left(
\frac{1}{(x^{13}{+}x^{23})^2}+\textrm{cyclic}\right)
+ \frac{\sfrac{5}{24}(m^1)^2}{(x^{12})^2+(x^{23})^2+(x^{31})^2}
+ \frac{\sfrac{1}{24}(m^0)^2}{(x^1{+}x^2{+}x^3)^2}\ .
\end{aligned}
\ee

Other translation and permutation invariant models may be constructed
by embedding the root systems of $I_2(3q)$ into three dimensions~\cite{glp2}.
The next higher case is $q=2$, i.e.~the $G_2$ model,
which is also obtained by combining the cases (\ref{h1}) and~(\ref{h2}).
The freedom of rescaling the short roots versus the long ones gives us
a more general solution,
\be
\begin{aligned}
F \= &-\sfrac14 f_{\rm{S}}\,(x^1{-}x^2)^2\ln|x^1{-}x^2|
\ -\ \sfrac{1}{12} f_{\rm{L}}\,(x^1{+}x^2{-}2x^3)^2\ln|x^1{+}x^2{-}2x^3|
\ +\ \textrm{cyclic}\\[4pt]
&-\sfrac12 f_R\,R^2\ln R
\ -\ \sfrac16 (x^1{+}x^2{+}x^3)^2\ln|x^1{+}x^2{+}x^3|
\qquad\textrm{with}\quad \sfrac32 f_{\rm S}+\sfrac32 f_{\rm L}+f_R=1\ .
\end{aligned}
\ee
The corresponding $u^A$ are determined by~(\ref{uxdef}), with
\be
h(\vp) \= h_0\,[\cos 3\vp ]^{1+f_{\rm S}}\
{}_2 F_1 \bigl( \sfrac{1+f_{\rm S}}{2}, \sfrac{1-f_{\rm L}}{2},
\sfrac{3+f_{\rm S}}{2}, \cos^2(3\vp) \bigr)\ .
\ee
For $f_{\rm L}=1$ or $f_{\rm S}=1$, this simplifies to
$h=h_0\,[\cos 3\vp]^{1+f_{\rm S}}$ or $h=h_0\,[\sin 3\vp]^{1+f_{\rm L}}$,
respectively. The `radial term' proportional to $R^2\ln R$ may be
eliminated in~$F$ by taking $f_R{=}0$ whence $f_{\rm L}{+}f_{\rm S}=\sfrac23$.
Thus,
\be
\begin{aligned}
(f_{\rm L},f_{\rm S})&=(1,-\sfrac13) \qquad\longrightarrow\qquad
h(\vp)\=h_0\,\cos^{2/3}(3\vp) \ ,\\[4pt]
(f_{\rm L},f_{\rm S})&=(-\sfrac13,1) \qquad\longrightarrow\qquad
h(\vp)\=h_0\,\sin^{2/3}(3\vp)\ .
\end{aligned}
\ee
These cases and the corresponding bosonic potential were already featured
in~\cite{glp1}.

\setcounter{equation}0
\section{Irreducible three-dimensional systems}

For irreducible systems beyond two dimensions, it is much more difficult to
solve the integrability condition (\ref{gw}) or~(\ref{simple}).
We again lower the inertial index and generalize~(\ref{uydef}) to
\be \label{3dans}
u^1 = y_1^2 + y_2^2 + y_3^2 =: R^2 \qquad\textrm{and}\qquad
u^2 = R^2\,h(\vt,\vp) \ ,\quad u^3 = R^2\,k(\vt,\vp) \ ,
\ee
where $\vt$ and $\vp$ are the two polar angles (declination and ascension) of
the two-sphere,
\be
y_1 \= R\,\sin\vt\,\cos\vp\ ,\qquad
y_2 \= R\,\sin\vt\,\sin\vp\ ,\qquad
y_3 \= R\,\cos\vt\ .
\ee
The matrix $(u^A_{\ i})$ is straightforwardly inverted.
Equating it to $(w_{A,i})^\top$ we discover that
\be
w_1 \= \sfrac12\ln R + \bw_1(\vt,\vp)  \qquad\textrm{and}\qquad
w_2 \= w_2(\vt,\vp) \ ,\quad w_3 \= w_3(\vt,\vp) \qquad\textrm{with}
\ee
\be
\begin{aligned}
\ph{\sfrac1{\sin^2\vt}}\pa_\vt\bw_1 &\= 
-\sfrac{h\,k_\vp-h_\vp k}{h_\vt k_\vp-h_\vp k_\vt}\ ,\qquad
\ph{\sfrac1{\sin^2\vt}}\pa_\vt w_2 \= 
\ \ \sfrac{k_\vp}{h_\vt k_\vp-h_\vp k_\vt}\ ,\qquad
\ph{\sfrac1{\sin^2\vt}}\pa_\vt w_3 \=\! 
-\sfrac{h_\vp}{h_\vt k_\vp-h_\vp k_\vt}\ ,\\[4pt]
\sfrac1{\sin^2\vt}\pa_\vp\bw_1 &\=
-\sfrac{h_\vt k-h\,k_\vt}{h_\vt k_\vp-h_\vp k_\vt}\ ,\qquad
\sfrac1{\sin^2\vt}\pa_\vp w_2 \=\! 
-\sfrac{k_\vt}{h_\vt k_\vp-h_\vp k_\vt}\ ,\qquad
\sfrac1{\sin^2\vt}\pa_\vp w_3 \=\, 
\ \sfrac{h_\vt}{h_\vt k_\vp-h_\vp k_\vt}\ ,
\end{aligned}
\ee
from which one learns that
\be
\pa_\vt\bw_1+h\,\pa_\vt w_2+k\,\pa_\vt w_3 \= 0 \qquad\textrm{and}\qquad
\pa_\vp\bw_1+h\,\pa_\vp w_2+k\,\pa_\vp w_3 \= 0 \ .
\ee
The corresponding superpotential reads
\be
G \= -\sfrac12 R^2\ln R - R^2\,\tg(\vt,\vp)
\qquad\textrm{with}\qquad
\tg \= \bw_1+h\,w_2+k\,w_3\ ,
\ee
leading to
\be
\tg_\vt \= h_\vt w_2 + k_\vt w_3 \qquad\textrm{and}\qquad
\tg_\vp \= h_\vp w_2 + k_\vp w_3\ .
\ee
Similarly, the analogous ansatz for~$F$ can be related to these functions,
and $V_B$ may be expressed through them as well, with $C=\sfrac12 m^1$.

In contrast to the $n{=}2$ case, the above equations do not admit solutions
for an arbitrary choice of $h(\vt,\vp)$ and $k(\vt,\vp)$. 
In fact, it seems quite nontrivial to find an admissible pair $(h,k)$ at all.
This is related to the appearance of the WDVV equations.
For completeness, we also display the integrability condition~(\ref{simple})
for our ansatz~(\ref{3dans}),
\be
\begin{aligned}
(k_\vp h_{\vp\vp}-h_\vp k_{\vp\vp})\ +\ 
(k_\vt h_{\vt\vp}-h_\vt k_{\vt\vp})\,\sin^2\vt &\=
(k_\vt h_\vp-h_\vt k_\vp)\,\sin 2\vt \ ,\\[4pt]
(k_\vp h_{\vp\vt}-h_\vp k_{\vp\vt})\ +\
(k_\vt h_{\vt\vt}-h_\vt k_{\vt\vt})\,\sin^2\vt &\= 0\ .
\end{aligned}
\ee

\subsection{$D_3$ solution}

Since some solutions for the prepotential~$F$,
based on Coxeter root systems~\cite{magra,ves}, are known,
we might as well take advantage of them and employ~(\ref{uwfromF})
to identify the inertial coordinates and superpotential for such cases.
Most important is the $A_3$ case, as it generalizes the four-particle 
Calogero model. We use the $D_3$ parametrization of the roots 
and allow for a `radial term' in the WDVV prepotential
\be \label{FD3}
F \= -\sfrac12 f_{\rm L} \sum_{i<j} (y_i{-}y_j)^2\ln|y_i{-}y_j|
-\sfrac12 f_{\rm L} \sum_{i<j} (y_i{+}y_j)^2\ln|y_i{+}y_j|
-\sfrac12 f_R R^2 \ln R
\ee
with $i,j=1,2,3$ and the restriction \ $4f_{\rm L}{+}f_R=1$.
For the special value \ $(f_{\rm L},f_R)=(-\sfrac14,2)$ \
we discovered the solution 
\be \label{D3sol}
u^1 \= R^2 \ ,\quad
u^2 \= R^2\,I(\sfrac{y_3}{r_1r_2}) ,\quad
u^3 \= R^2\,I(\sfrac{r_2}{r_1}) \qquad\textrm{with}\qquad
r_i^2 \= y_i+\sqrt{y_i^2{-}y_3^2}\ ,
\ee
where \ $I(x)=\int_0^x \sfrac{\di t}{\sqrt{1-t^4}}$ \
denotes an incomplete elliptical integral of the first kind.
The inverse Jacobian yields
\be
\begin{aligned}
w_{2,1} &\= \frac{r_1r_2}{y_3R^4}\,
\sqrt{\sfrac{2(y_1^2-y_2^2)}{y_1r_1^2-y_2r_2^2}}\,
\biggl( y_1y_2\sqrt{y_2^2{-}y_3^2}-(y_2^2{+}y_3^2)\sqrt{y_1^2{-}y_3^2}\biggr)
\ ,\\
w_{2,2} &\= \frac{r_1r_2}{y_3R^4}\, 
\sqrt{\sfrac{2(y_1^2-y_2^2)}{y_1r_1^2-y_2r_2^2}}\,
\biggl( y_1y_2\sqrt{y_1^2{-}y_3^2}-(y_1^2{+}y_3^2)\sqrt{y_2^2{-}y_3^2}\biggr)
\ ,\\
w_{2,3} &\= \frac{r_1r_2}{R^4}\,
\sqrt{\sfrac{2(y_1^2-y_2^2)}{y_1r_1^2-y_2r_2^2}}\,
\biggl( y_1\sqrt{y_1^2{-}y_3^2}+y_2\sqrt{y_2^2{-}y_3^2} \biggr)
\ ,\\
w_{3,1} &\= \frac{r_2}{r_1^3R^4}\,
\sqrt{2(y_1r_1^2{-}y_2r_2^2)}\,
\Bigl(r_1^2r_2^2y_1-3r_1^2y_1y_2-r_2^2(y_2^2{+}y_3^2)+2y_2(y_2^2{+}y_3^2)\Bigr)
\ ,\\
w_{3,2} &\= \frac{r_2}{r_1^3R^4}\,
\sqrt{2(y_1r_1^2{-}y_2r_2^2)}\,
\Bigl(r_1^2r_2^2y_2+r_1^2(y_1^2{-}2y_2^2{+}y_3^2)+r_2^2y_1y_2-2y_1y_2^2\Bigr)
\ ,\\
w_{3,3} &\= \frac{r_2}{r_1^3R^4}\,
\sqrt{2(y_1r_1^2{-}y_2r_2^2)}\,
\Bigl(-r_1^2r_2^2(y_1^2{+}y_2^2)+r_1^2y_2(2y_1^2{+}2y_2^2{-}y_3^2)+
       r_2^2y_1y_3^2-2y_1y_2y_3^2\Bigr)
\end{aligned}
\ee
and $w_{1,i}$ in terms of elliptic integrals,
which gives us $U$ and the bosonic potential
\be
\begin{aligned}
V_B \=&-\frac2{(R^2)^3}\biggl\{ \left((m^2)^2{+}(m^3)^2\right)
\left[y_1\,(y_2^2{-}y_3^2)^{\frac32} + y_2\,(y_1^2{-}y_3^2)^{\frac32}\right]
\\ &+
\sqrt{2}\,m^2m^3\frac{(y_1^2{-}y_2^2)^{\frac32}}{y_3}\sqrt{r_1^2{+}r_2^2}\,
r_2\,(r_2^2{-}2y_2)\sqrt{\frac{y_3^2+r_2^2(r_1^2{-}2y_2)}{y_1r_1^2-y_2r_2^2}}
\biggr\} \ +\ \textrm{$m^1$-terms}\ .
\end{aligned}
\ee
It is regular except for $R\to0$.

To pass to the $A_3$ parametrization $(z_1,z_2,z_3)$, one has to apply
the orthogonal transformation
\be
y_i \= O_{ij}\,z_j \qquad\textrm{with}\qquad
\bigl(O_{ij}\bigr) \= \sfrac{1}{\sqrt{6}}\,\biggl(\!\!\begin{smallmatrix}
\ph{-}\sqrt{3} & \ \ph{-}1 & \ -\sqrt{2} \\ 
-\sqrt{3} & \ \ph{-}1 & \ -\sqrt{2} \\
\ph{-}0 & \ -2 & \ -\sqrt{2}
\end{smallmatrix}\biggr)\ ,
\ee
so that the six positive roots become
\be
\sfrac{2\sqrt{3}z_1}{\sqrt{6}}\ ,\quad
\sfrac{\sqrt{3}z_1+3z_2}{\sqrt{6}}\ ,\quad
\sfrac{-\sqrt{3}z_1+3z_2}{\sqrt{6}}\ ,\quad
\sfrac{2z_2-2\sqrt{2}z_3}{\sqrt{6}}\ ,\quad
\sfrac{\sqrt{3}z_1-z_2-2\sqrt{2}z_3}{\sqrt{6}}\ ,\quad
\sfrac{-\sqrt{3}z_1-z_2-2\sqrt{2}z_3}{\sqrt{6}}\ .
\ee

\subsection{$B_3$ solution}

Surprisingly, a simpler solution arises for the $B_3$ root system, 
with the WDVV~prepotential
\be \label{FB3}
F \= -\sfrac12 f_{\rm S} \sum_i y_i^2\ln|y_i|
-\sfrac12 f_{\rm L} \sum_{i<j} (y_i{-}y_j)^2\ln|y_i{-}y_j|
-\sfrac12 f_{\rm L} \sum_{i<j} (y_i{+}y_j)^2\ln|y_i{+}y_j|
\ee
lacking a radial term. The weights are constrained by
$f_{\rm S}+4f_{\rm L}=1$. For $(f_{\rm S},f_{\rm L})=(5,-1)$ we found the
inertial coordinates
\be \label{b3sol}
u^1 \= \frac{y_1^6}{(y_1^2{-}y_2^2)(y_1^2{-}y_3^2)}\ ,\quad
u^2 \= \frac{y_2^6}{(y_2^2{-}y_3^2)(y_2^2{-}y_1^2)}\ ,\quad
u^3 \= \frac{y_3^6}{(y_3^2{-}y_1^2)(y_3^2{-}y_2^2)}\ ,
\ee
which yield the dual coordinates
\be
w_1 \= \sfrac12\ln|y_1| +
\sfrac{y_2^2+y_3^2}{8\,y_1^2} - \sfrac{y_2^2 y_3^2}{24\,y_1^4}\ ,\quad
w_2 \= \sfrac12\ln|y_2| +
\sfrac{y_3^2+y_1^2}{8\,y_2^2} - \sfrac{y_3^2 y_1^2}{24\,y_2^4}\ ,\quad
w_3 \= \sfrac12\ln|y_3| +
\sfrac{y_1^2+y_2^2}{8\,y_3^2} - \sfrac{y_1^2 y_2^2}{24\,y_3^4}\ .
\ee
Note that this solution is outside the ansatz~(\ref{3dans})
and somewhat peculiar since $w_A$ contains rational parts but features
logarithms of the short roots only.
Moreover, it is invariant under permutations of the~$y_i$ but of course
not translation invariant. No coordinate is distinguished as radial,
but we have \ $u^1{+}u^2{+}u^3=y_1^2{+}y_2^2{+}y_3^2=R^2$. Due to $c_A=1/2$,
the central charge becomes $C=\sfrac12(m^1{+}m^2{+}m^3)$.
We read off the second prepotential
\be
U \= -\sfrac12 m^1 \ln|y_1| - m^1\left(
\sfrac{y_2^2+y_3^2}{8\,y_1^2} - \sfrac{y_2^2 y_3^2}{24\,y_1^4} \right)
\ +\ \textrm{cyclic}\ ,
\ee
thus obtaining the specific homogeneous function 
needed to overcome the $n{=}3$ barrier of~\cite{glp2}.
It displays the expected singular behavior
$U\sim|y_i|^{1-f_{\rm S}}$ for $y_i\to0$
and has couplings only for the short roots.
The rational parts of $w_A$ drop out in the superpotential 
\be
\begin{aligned}
G &\= \frac{y_1^6\,\ln|y_1|}{2(y_1^2{-}y_2^2)(y_3^2{-}y_1^2)}\ +\
\frac{y_2^6\,\ln|y_2|}{2(y_2^2{-}y_3^2)(y_1^2{-}y_2^2)}\ +\
\frac{y_3^6\,\ln|y_3|}{2(y_3^2{-}y_1^2)(y_2^2{-}y_3^2)}\\[4pt]
&\= -\sfrac12 u^1\,\ln|y_1(u)| 
-\sfrac12 u^2\,\ln|y_2(u)| -\sfrac12 u^3\,\ln|y_3(u)|\ ,
\end{aligned}
\ee
but we could not invert (\ref{b3sol}) to obtain $y_i(u)$. 
Finally, one may compute the bosonic potential
\be
V_B \= \sfrac1{288}\left[
m^1\Bigl(\sfrac{y_2^2}{y_1^4}-\sfrac{3}{y_1^2}\Bigr)y_3 +
m^2\Bigl(\sfrac{y_1^2}{y_2^4}-\sfrac{3}{y_2^2}\Bigr)y_3 -
m^3\Bigl(2\sfrac{y_1^2y_2^2}{y_3^5}-3\sfrac{y_1^2+y_2^2}{y_3^3}
                                   +\sfrac{6}{y_3}\Bigr)
\right]^2 \ +\ \textrm{cyclic}\ ,
\ee
which features poles (up to tenth order) for the short roots only.
Deviating from the above special values of $(f_{\rm L},f_{\rm S})$
destroys the simplicity of this solution.

\setcounter{equation}0
\section{Embedding into four dimensions}

We may try to produce a translation-invariant four-particle model by repeating
the previous story one dimension higher. To this end, we employ the embedding
\be \label{3to4}
y^i \= M^i_{\ \mu}\,x^\mu \qquad\textrm{with}\qquad
\bigl(M^i_{\ \mu}\bigr) \= \begin{pmatrix}
\sfrac{1}{\sqrt{2}} &      -\sfrac{1}{\sqrt{2}} & \ph{-}0 & \ph{-}0 \\[6pt]
\sfrac{1}{\sqrt{6}} & \ph{-}\sfrac{1}{\sqrt{6}} & -\sfrac{2}{\sqrt{6}} &
\ph{-} 0 \\[6pt]
\sfrac{1}{\sqrt{12}}& \ph{-}\sfrac{1}{\sqrt{12}}& \ph{-}\sfrac{1}{\sqrt{12}} &
-\sfrac{3}{\sqrt{12}} \end{pmatrix}\ .
\ee
where the partial isometry $M$ maps onto the relative-motion space due to
\be
M\,M^\top \= \unity_3 \qquad\textrm{and}\qquad
M^\top M \= P \= \sfrac14\,\biggl(\!\!\begin{smallmatrix}
\ph{-}3 & -1 & -1 & -1 \\ -1 & \ph{-}3 & -1 & -1 \\
-1 & -1 & \ph{-}3 & -1 \\ -1 & -1 & -1 & \ph{-}3
\end{smallmatrix}\biggr)\ .
\ee
For embedding our $D_3$ solution as an $A_3$ model, we must apply the map~$M$
to the $z^i$~coordinates, i.e.
\be \label{3to4a}
y^i \= (O\,M)^i_{\ \mu}\,x^\mu \qquad\textrm{with}\qquad
\bigl((O\,M)^i_{\ \mu}\bigr) \= \sfrac12\,\biggl(\!\!\begin{smallmatrix}
\ph{-}1 & -1 & -1 & \ph{-}1 \\ 
-1 & \ph{-}1 & -1 & \ph{-}1 \\
-1 & -1 & \ph{-}1 & \ph{-}1
\end{smallmatrix}\biggr)\ .
\ee
Together with the center of mass \ $y^0=\sfrac12(x^1{+}x^2{+}x^3{+}x^4)$,
this is the triality map relating $D_4$ vectors to spinors.
The center-of-mass degree of freedom is decoupled,
\be
u^0 \= (x^1{+}x^2{+}x^3{+}x^4)^2 \qquad\longrightarrow\qquad
w_0 \= \sfrac18\ln\left|x^1{+}x^2{+}x^3{+}x^4\right|
\= \sfrac{1}{16}\ln u^0\ .
\ee
For the relative motion, our ansatz~(\ref{3dans}) extends to
\be \label{4dans}
\begin{aligned}
u^1\=\tR^2\ ,\qquad& u^2\=\tR^2\,h(\vt,\vp)\ ,\qquad u^3\=\tR^2\,k(\vt,\vp)
\\[4pt] \textrm{with}\qquad \tR^2=\sfrac14\sum_{i<j}(x^{ij})^2
\qquad\textrm{and}\qquad& (y^1,\,y^2,\,y^3) \= 
\tR\,(\sin\vt\,\cos\vp\,,\,\sin\vt\,\sin\vp\,,\,\cos\vt)\ .
\end{aligned}
\ee

Models of identical particles require 
invariance under permutations of the $x^i$ coordinates.
The permutation group ${\cal S}_4$ acts on the two-sphere~$(\vt,\vp)$ as the 
Weyl group of~$A_3$, i.e.~by permuting the corners of a regular tetrahedron
by via $\frac{2\pi}{3}$ rotations and reflections.
Therefore, a permutation-invariant solution requires $h$ and $k$ to be
${\cal S}_4$ invariant functions. Such functions are generated by
taking some function $\tilde{h}(\vt,\vp)$ and forming a symmetric
combination from its pullbacks $(\tilde{h}\circ\pi)(\vt,\vp)$ 
along the ${\cal S}_4$ orbit. 
The simplest option just averages~(\ref{4dans}) over its ${\cal S}_4$ orbit.
This is admissible due to the linearity of~(\ref{uwfromF}) 
(assuming a permutation symmetric~$F$ is given),
but may result in a degenerate solution. 
In this way, our $D_3$ solution~(\ref{D3sol}), after embedding into 
four dimensions via $y{=}OMx$ and averaging over ${\cal S}_4$ permutations, 
may yield a totally symmetric four-particle system after all, 
although we have not checked this.

Another four-particle model is created by
subjecting our $B_3$ solution to the embedding~(\ref{3to4}), 
Clearly, the corresponding four-dimensional superpotential~$G(x)$ and 
bosonic potential~$V_B(x)$ are not invariant under permutations of the~$x^i$. 
This is hardly surprising, since this system started out being only 
${\cal S}_3$ symmetric, and so an ${\cal S}_4$ average of the above solution 
is not consistent with the WDVV solution~(\ref{FB3}).

In order to produce a genuine four-particle ${\cal N}{=}4$ Calogero system,
one has to find a solution which combines the features of our $D_3$ and
$B_3$ systems above, namely for 
\be \label{FA3}
(f_{\rm S},f_{\rm L},f_R) \= (0,\sfrac14,0) \qquad\longleftrightarrow\qquad
F \= 
-\sfrac18 \sum_{i<j} (y^i{-}y^j)^2\ln|y^i{-}y^j|
-\sfrac18 \sum_{i<j} (y^i{+}y^j)^2\ln|y^i{+}y^j|\ .
\ee
We know~\cite{glp2} that $U$ (and therefore some $w_A$) behaves as 
$|\ay|^{1-f_\alpha\alpha{\cdot}\alpha}$ when crossing the wall $\ay{=}0$
for any root~$\alpha$, thus no logarithms should occur in
\be
U(y)\ \sim\ (y^i{\mp}y^j)^{1/2} \quad\textrm{for}\quad y^i\to\pm y^j
\qquad\textrm{hence}\qquad
U(x)\ \sim\ (x^i{-}x^j)^{1/2} \quad\textrm{for}\quad x^i\to x^j\ .
\ee
It remains a challenge to construct the superpotential~$G$ and prepotential~$U$
belonging to~(\ref{FA3}).

\section*{Acknowledgements}

We are indebted to Anton Galajinsky and Anton Sutulin for collaboration
at an early stage of this project. 
S.K.\ is grateful to A.P.~Isaev for discussions.
K.P.\ thanks the ITP at Leibniz Universit\"at Hannover for hospitality
and the DAAD and the Dynasty Foundation for support.
\vskip .2cm
This work was partially supported by INTAS under contract 05-7928,
by an RF Presidential grant NS-2553.2008.2 as well as by the grants
RFBR-08-02-90490-Ukr, 06-02-16684, 06-01-00627-a, and  DFG~436 Rus~113/669/03.

\end{document}